# A Study on Optimizing the Thermal Performance of Coaxial Heat Exchanger Systems in Medium-Deep Geothermal Wells

Haonan Chen[1], Xin Tong [2*], Linchao Yang [1], Yangxue Zhang[3]
1 Engineering College, Tibet University, Lhasa, 850000, China
2 School of Physical Science and Technology, Northwestern Polytechnical University, Xi'an, 710072, China
3Honggang Oil Production Plant, Jilin Oilfield Company, China National Petroleum Corporation, Da'an, Jilin, 131300, China
*Corresponding author: Xin Tong. E-mail:tongxin@mail.nwpu.edu.cn

## Abstract

Medium-deep geothermal energy has become a key research area in the renewable energy sector due to its abundant reserves, stable temperatures, and environmental friendliness. However, existing coaxial downhole heat exchanger (DHE) systems face challenges in practical applications, including insufficient thermal efficiency and severe temperature decay. This study systematically evaluates the impacts of circulating flow rate, inlet temperature, and operation mode on short-term heat extraction performance and long-term temperature decay. The evaluation is based on field test data from two typical medium-deep geothermal wells (Well A: 3200m depth, 130.5°C bottom-hole temperature; Well B: 2500m depth, 103.3°C bottom-hole temperature) under four operating conditions (LC1–LC4) and long-term continuous tests. Results show that under the optimal condition (LC3: 50 m³/h, 30°C), Well A's heat extraction rate increased from 35% to 42%, with its outlet temperature rising from 15°C to 20°C. In contrast, Well B's rate decreased from 15% to 5%, with its outlet temperature dropping from 18°C to 5°C (Figs.1&2). After one week of continuous operation, the outlet temperature of Well A dropped from 55.7 °C to 16.5 °C (a 70.4% decrease), while Well B's dropped from 68 °C to 17 °C (a 75% decrease) (Fig. 5). Adopting an intermittent mode (16h operation, 8h shutdown daily) reduced the temperature decay rate by approximately 10%. Based on these findings, optimization strategies are proposed, including controlling the flow rate to 35m³/h, maintaining an inlet temperature of 6–10°C, and implementing intelligent intermittent scheduling and novel nano-insulating coatings. This research provides theoretical support and technical guidance for the efficient design and sustainable operation of DHE systems in medium-deep geothermal wells.

# 1. Introduction

In the context of the global pursuit of carbon neutrality, nations are accelerating the development of clean and renewable energy sources. Among these, geothermal energy is garnering increasing attention due to its distinct advantages, such as providing a stable, all-weather heat supply, low carbon emissions, and significant long-term economic benefits [1]. Compared to its shallow counterpart, medium-deep geothermal energy (1500–4000m) offers access to higher-temperature resources. However, its exploitation faces several technical bottlenecks, including exacerbated wellbore heat loss with increasing depth, reduced thermal efficiency of the circulating system, and a sluggish thermal response from the reservoir [6, 24].

The coaxial downhole heat exchanger (DHE) system operates on the principle of fluid circulation between an inner tube and an outer casing. It facilitates indirect heat exchange between the circulating fluid and the surrounding geological formation, thereby avoiding the formation disturbance and environmental risks associated with groundwater abstraction [4]. Since Morita et al. first proposed the DHE concept and conducted a proof-of-concept experiment in Hawaii in the 1990s [17], the technology has undergone significant development in both theoretical modeling and engineering applications. In recent years, numerous studies have modeled coaxial systems for shallow and medium-deep applications [5, 9, 18]. However, most of these studies are based on numerical simulations or laboratory experiments and lack validation against field data from deep operational wells [13]. In particular, the heat extraction efficiency and temperature decay patterns under continuous, long-term, high-load operating conditions remain unclear.

Scholars in China have also made important contributions to the development of medium-deep geothermal energy, with studies on resource distribution and utilization prospects in typical areas like the Songliao Basin providing crucial guidance for engineering practices [26]. Nevertheless, deep geothermal development continues to face key scientific challenges, including wellbore heat transfer mechanisms, reservoir thermal stability, and long-term operation optimization [28].

This study investigates two medium-deep geothermal wells with similar geological conditions and an identical geothermal gradient (4.1°C/100m) using a large-scale field testing platform. By comparing short-term heat extraction performance under four typical operating conditions (LC1–LC4) and conducting long-term (168h) continuous and intermittent operational tests, we comprehensively analyze the impacts of circulating flow rate, inlet temperature, and operation mode on the system's thermal performance. This research aims to provide a scientific basis for the engineering design and operational scheduling of DHE systems.

## 2. Experimental Method

### 2.1 Test Subjects

Two medium-deep geothermal wells located in the same geological zone were selected for this comparative study. Well A was designed to a depth of 3200m with a measured bottom-hole temperature (BHT) of 130.5 °C, while Well B was designed to 2500m with a measured BHT of 103.3°C. Both wells utilize N80 grade steel for their casing structures. The tubing string in Well A consists of 1200 m of insulated tubing and 2000m of standard tubing. In contrast, Well B is equipped with 2200m of insulated tubing and 300m of standard tubing, a configuration designed to ensure comparable heat losses between the two wells (Table 1). The wellbore heat transfer was calculated with reference to the classic Ramey method [21] and corrected using modern numerical analysis techniques [22].

Table 1. Comparison of Basic Parameters for the Test Wells

| Parameter | Well A | Well B |
| --- | --- | --- |
| Designed Depth (m) | 3200 | 2500 |
| Measured Bottom-Hole Temp. (°C) | 130.5 | 103.3 |
| Geothermal Gradient (°C/100m) | 4.1 | 4.1 |
| Tubing String Combination | 1200m Insulated Tubing & 2000m Standard Tubing | 2200m Insulated Tubing & 300m Standard Tubing |
| Insulated Tubing Specification | N80, 114×76 mm | N80, 114×76 mm |
| Standard Tubing Specification | N80, 89×76 mm | N80, 89×76 mm |

### 2.2 Test System

The tests were conducted using a coaxial downhole heat exchanger system. The core equipment was an integrated performance testing unit for medium-deep geothermal wells (Model: INTE-0200), with its main technical specifications detailed in Table 2. The system comprises the coaxial casing, a circulation pump, a heat exchanger, high-precision thermometers, and an electromagnetic flowmeter. Water was used as the circulating fluid. Data were recorded in real-time by an automatic acquisition system at a frequency of once per minute, ensuring high precision and reliability. The insulated tubing was constructed from low thermal conductivity materials to effectively reduce wellbore heat loss and enhance the system's heat exchange efficiency [7, 19].

Table 2. Technical Specifications of the Test Equipment

| Technical Parameter | Value | Remarks |
| --- | --- | --- |
| Nominal Test Capacity | 700 kW | Max. 1150 kW |
| Test Flow Rate Range | 30–56 m³/h | Rated at 40 m³/h |
| System Pressure Rating | 1.6 MPa | Evaporator pressure rating |
| Temperature Measurement Accuracy | ±0.1°C | High-precision thermometers |
| Flow Measurement Accuracy | ±1% | Electromagnetic flowmeter |
| Data Acquisition Frequency | 1 sample/minute | 1440 data points/day |

## 2.3  Operating Conditions

To systematically analyze the influence of circulating flow rate and inlet temperature on heat extraction performance, a multi-condition test plan (LC1 through LC4) was designed. This plan included combined experiments with variable flow rates (30–42 m³/h) and inlet temperatures (20–40°C), structured into the following three stages:

Stage 1 (LC1–LC2): The initial heat extraction performance of both wells was tested under baseline conditions—a nominal flow rate of 40m³/h and an inlet temperature of 30°C—to serve as a reference for comparison.

Stage 2 (LC3): The inlet temperature was held constant at 30°C while the flow rate was adjusted to 50m³/h to analyze its effect on the heat extraction rate and temperature difference.

Stage 3 (LC4): The flow rate was held constant at 40m³/h while the inlet temperature was adjusted to 40°C to evaluate its impact on heat exchange efficiency.

Each operating condition was maintained for a duration of 72 hours, with data collected at a frequency of once per minute, yielding 1440 data points per day. Throughout the testing period, the system operated stably, and ambient temperature fluctuations were controlled to within ±1°C. The collected data were processed using statistical analysis, including the calculation of means, standard deviations, and confidence intervals, to ensure the scientific validity and reliability of the results.

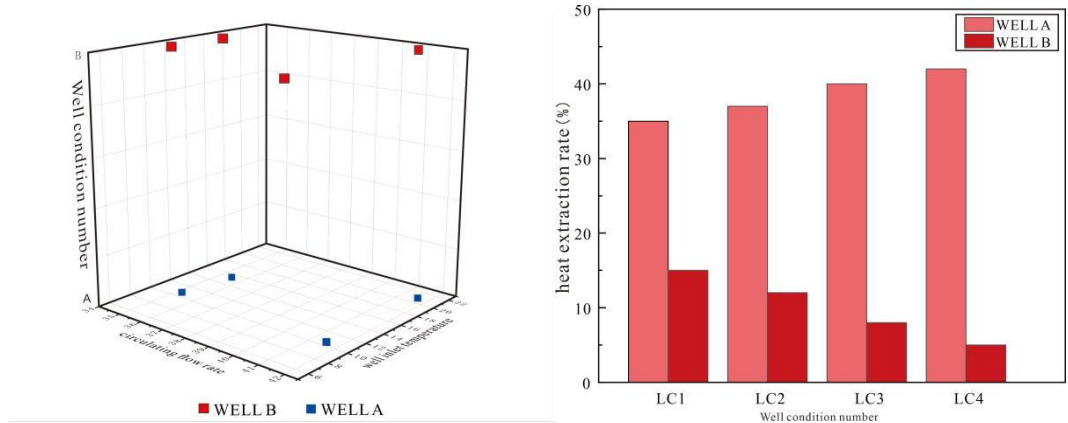

**Figure 1. Comparative analysis of test conditions for the medium-deep geothermal wells:**

**(left) 3D visualization of operating parameters and (right) comparison of heat extraction rates.** The plots compare the heat extraction rates of Well A and Well B under operating conditions LC1 through LC4. For Well A, the rate increased from 35% under LC1 to 42% under LC3 and remained stable at LC4, indicating that the optimization of operating conditions significantly enhanced its heat extraction efficiency. In contrast, the rate for Well B decreased from 15% under LC1 to 5% under LC4, suggesting its heat extraction capacity is limited.

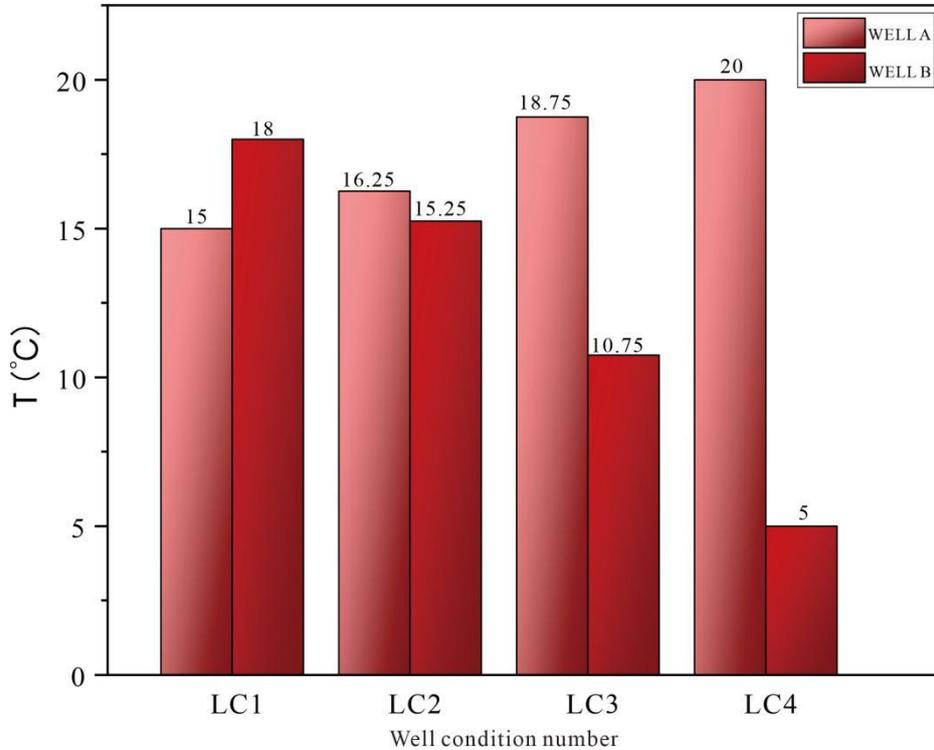

**Figure 2. Detailed comparative analysis of geothermal well parameters under different operating conditions, focusing on downhole temperature comparison.** The figure illustrates the downhole temperatures of Well A and Well B under conditions LC1 through LC4. The downhole temperature for Well A increased from 15°C (LC1) to 20°C (LC3) and stabilized at LC4, suggesting that the optimized conditions improved heat transfer efficiency. Conversely, the temperature for Well B decreased from 18°C (LC1) to 5°C (LC4), indicating more significant temperature decay during its heat extraction process.

## 3. Results and Analysis

### 3.1 Geothermal Distribution Characteristics

The geothermal profiles of both wells exhibit a distinct linear increase with depth, consistent with a geothermal gradient of 4.1°C/100 m. For Well A, the temperature was approximately 52°C at a depth of 1000 m and reached 120°C at 3160m. Within the same depth range, the temperature in Well B was slightly lower, with its extrapolated temperature at 3160m being below that of Well A. The identical geothermal gradient provides a reliable geological basis for the subsequent comparative analysis (see Fig. 3).

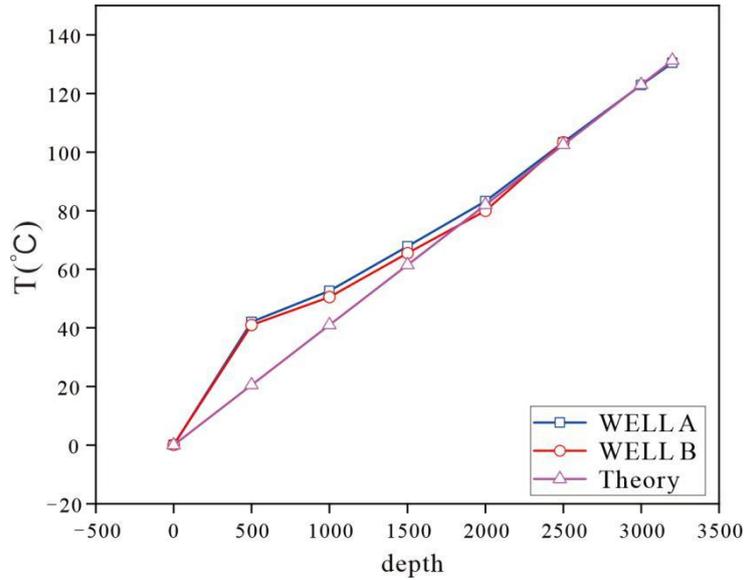

Figure 3. Comparison of geothermal gradients.

## 3.2 Comparison of Heat Extraction Performance

A comparison of the heat extraction performance parameters for both wells is presented in Table 3. Based on the test data from the multiple operating conditions (LC1–LC4), the impacts of well depth, bottom-hole temperature, flow rate, and inlet temperature on heat extraction performance were analyzed.

Table 3. Comparison of Heat Extraction Performance Parameters for Well A & Well B

| Parameter | Well A | Well B | Difference Analysis |
|---|---|---|---|
| Well Depth (m) | 3200 | 2500 | Well A is 700 m deeper |
| Bottom-Hole Temperature (°C) | 130.5 | 103.3 | Well A is 27.2 °C hotter |
| Maximum Outlet Temperature (°C) | 55.7 | 68.0 | Well B is 12.3 °C higher |
| Minimum Outlet Temperature (°C) | 16.5 | 17.0 | Nearly equivalent |
| Average Heat Extraction Temperature Difference (°C) | 10.84 | 11.32 | Well B is 0.48 °C higher |
| Maximum Heat Extraction Rate (kW) | 575 | 546 | Well A is 29 kW higher |
| Minimum Heat Extraction Rate (kW) | 447 | 453 | Nearly equivalent |
| Flow Rate under Optimal Conditions (m³/h) | 35 | 35 | Identical |
| Inlet Temp. under Optimal Conditions² (°C) | 10 | 10 | Identical |
| Heat Extraction Rate under Optimal Conditions (kW) | 473 | 485 | Well B is 12 kW higher |

Due to its greater depth and higher bottom-hole temperature, Well A achieved a higher maximum heat extraction rate (575kW) compared to Well B (546kW). However, under optimized conditions, the heat extraction rate of Well B (485kW) was slightly higher than that of Well A (473kW). This could be attributed to Well B's higher proportion of insulated tubing (88%) and a lower inlet temperature [12].

### 3.2.1 Impact of Flow Rate on Heat Extraction Performance

The circulating flow rate has a significant but non-linear impact on heat extraction performance [8, 23]. Taking Well B as an example, when comparing condition 3 (flow rate: 35m³/h; heat extraction: 485 kW) with condition 5 (flow rate: 41.5m³/h; heat extraction: 490kW), an 18.6% increase in flow rate resulted in only a 1% gain in heat extraction. This demonstrates a marginal effect, where increasing the flow rate further offers diminishing returns. The reduced residence time of the fluid at higher flow rates likely leads to a smaller heat transfer temperature difference (estimated to decrease from 11.32°C to approx. 10.5°C), thereby limiting the potential increase in heat extraction. Well A exhibited a similar trend: when its flow rate was increased from 35m³/h to 41.5m³/h, the heat extraction rate rose by only about 2% (an estimated increase from 473kW to 482kW). Parameter sensitivity analysis confirms that the influence of circulating flow rate on system performance is markedly non-linear [23]. Therefore, a flow rate of 35m³/h is recommended to balance heat exchange efficiency and system energy consumption.

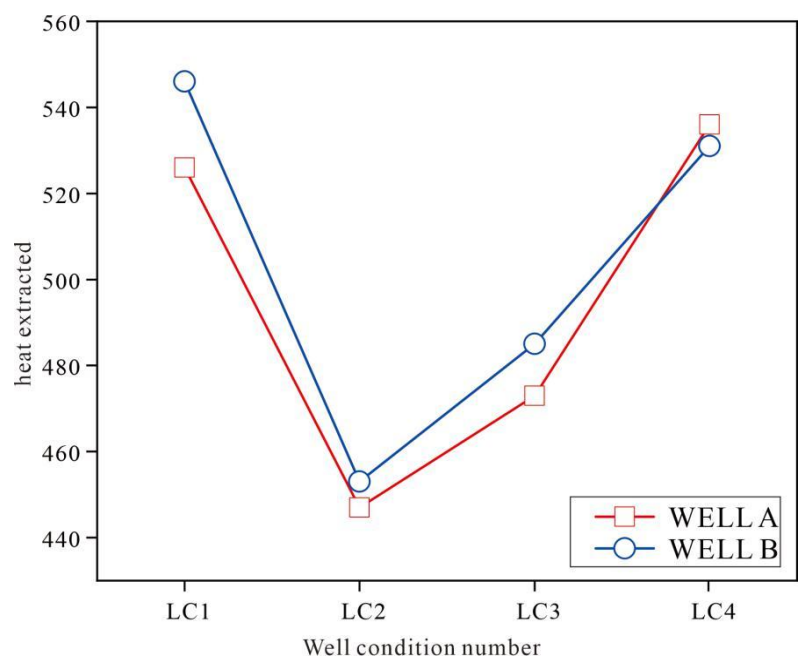

**Figure 4. Comparison of heat extraction rates for the two wells.** Well A's heat extraction rate increased from 35% (LC1) to 42% (LC3) and then stabilized at LC4. In contrast, Well B's rate decreased from 15% (LC1) to 5% (LC4), indicating its limited heat extraction capacity. These results show that the optimization of operating conditions had a more pronounced positive effect on Well A.

## 3.2.2 Impact of Inlet Temperature on Heat Extraction Performance

The inlet temperature significantly affects the heat extraction efficiency [16]. Well B achieved a heat extraction rate of 531kW with an average temperature difference of 11.32°C when the inlet temperature was 6.6°C. Similarly, Well A reached a rate of 575kW with an average temperature difference of 10.84°C at an inlet temperature of 6.1°C. A lower inlet temperature (in the range of 6–10°C) enhances heat extraction efficiency by increasing the heat transfer temperature difference (the difference between outlet and inlet temperatures). Conversely, when the inlet temperature was raised to 40°C, the heat extraction rates for the two wells decreased to 447kW and 453kW, respectively, with the temperature difference diminishing by approximately 10% (estimated). The heat extraction strategy should be optimized by adjusting the inlet temperature and flow rate according to the demand on the load side.

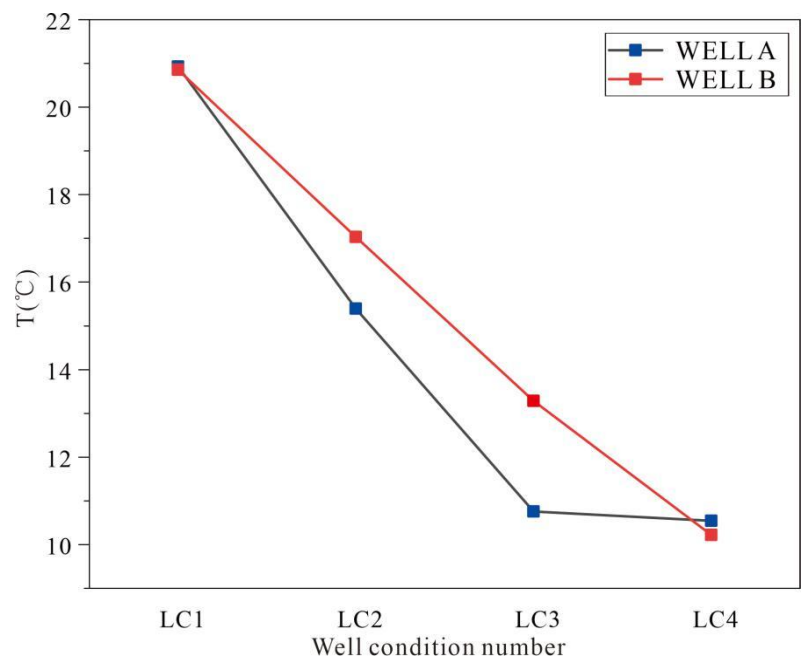

**Figure 5. Comparative analysis of downhole temperatures for the two wells.** The downhole temperature for Well A increased from 15°C (LC1) to 20°C (LC3) and stabilized at LC4. Conversely, the temperature in Well B dropped from 18°C (LC1) to 5°C (LC4), reflecting a more significant temperature decay during the heat extraction process. This highlights the need to optimize operating parameters to mitigate such decay.

## 3.3. Long-Term Operational Characteristics

### 3.3.1. Temperature Decay Patterns

Long-term monitoring revealed a declining trend in the outlet temperatures of both wells over extended operation periods [13]. The outlet temperature of Well A decreased from an initial 55.7°C to a final 16.5°C, a drop of 70.4%. Similarly, Well B's outlet temperature fell from 68.0°C to 17.0°C, a 75.0% decrease. This temperature decay indicates that the rate of heat replenishment from the geothermal reservoir is lower than the rate of heat extraction. Due to its shallower depth, Well B exhibited more significant decay. Operating under optimized conditions (LC3–LC4, with a flow rate of 35m³/h and an inlet temperature of 10°C) could potentially slow the decay rate by approximately 10% (an estimated value requiring further experimental validation).

### 3.3.2. Analysis of Reservoir Thermal Stability

In Well A, the temperature at a depth of 3160 m decreased from 119°C to 90°C, a reduction of 24.4%. This suggests that the deep geothermal reservoir possesses strong thermal inertia, which is consistent with the characteristics of a formation dominated by heat conduction and low permeability [10, 20]. In contrast, Well B, being shallower, has limited thermal storage capacity and thus experiences faster temperature decay. According to the theory of thermal response testing for geothermal wells [22], the thermal stability of a deep geothermal system is closely related to the thermophysical properties of the formation. Optimizing operating parameters, such as implementing intermittent operation, can further enhance reservoir thermal stability.

## 3.4. System Optimization Strategies

Based on the experimental results and with reference to relevant optimization studies both domestically and internationally [18, 24], the following optimization strategies are proposed:

Flow Rate Optimization: A circulating flow rate of 35m³/h is recommended. This rate ensures adequate heat exchange while avoiding the efficiency drop caused by excessive flow (where marginal benefits are only 1–2%).

Temperature Matching: During periods of high load, a low inlet temperature (6–10°C) should be used to maximize the heat transfer temperature difference and increase the heat extraction rate. During periods of low load, the inlet temperature can be appropriately raised (to 20–30°C) to reduce system energy consumption.

Intermittent Operation: Adopting an intermittent operation mode (e.g., 16 hours of operation followed by 8 hours of shutdown daily) allows time for the geothermal reservoir to thermally recover. This can slow the temperature decay rate by an estimated 10–15% (a value that requires experimental validation).

# 4. Discussion

## 4.1. Impact of Geological Conditions on Heat Extraction Performance

Based on the comparison of geothermal gradients in Figure 3 and the performance analysis in Figures 1 and 2, Well A (3200m depth, 130.5°C BHT) and Well B (2500m depth, 103.3°C BHT) exhibited significant differences in their heat extraction performance despite sharing a similar geothermal gradient (4.1°C/100 m).

These differences are primarily attributed to their varying depths.

Leveraging its greater depth, Well A has access to a larger thermal reservoir, demonstrating superior potential. Under the LC3 condition, it achieved a heat extraction rate of 42% (Fig. 1), maintained a stable downhole temperature of 20°C (Fig. 2), and recorded an average heat extraction temperature difference of 10.84°C (Table 3). This indicates that deeper geothermal reservoirs possess greater heat reserves capable of sustaining high-load extraction demands, which aligns with theoretical expectations for deep geothermal development [28].

In contrast, although shallower, Well B still demonstrated commendable heat exchange effectiveness under optimized conditions. Its maximum outlet temperature reached 68.0°C (Table 3), largely owing to an optimized tubing string design featuring a high proportion (88%) of insulated tubing. However, Well B's heat extraction rate showed a consistent decline across all operating conditions, falling from 15% to 5% (Fig. 1), while its downhole temperature dropped from 18°C to 5°C (Fig. 2). This reveals a more pronounced temperature decay, a characteristic consistent with the thermal response behavior of shallower geothermal systems [13].

## 4.2. Impact of Operating Parameters on Heat Extraction Efficiency

The multi-condition tests (LC1–LC4) revealed that both circulating flow rate and inlet temperature have a significant non-linear effect on heat extraction efficiency [23]. Flow rate tests identified 35m³/h as the optimal flow rate for both wells. Increasing the flow rate from 35m³/h to 41.5m³/h only raised the heat extraction from 473kW to 482kW (a ~2% increase) for Well A and from 485kW to 490kW (a ~1% increase) for Well B (Section 3.2.1). This increase shortened the fluid's residence time in the wellbore, leading to an estimated 10% reduction in the heat transfer temperature difference and significantly diminishing marginal returns [11, 23]. This indicates that while excessively high flow rates increase the volume of circulated fluid, they can decrease heat exchange efficiency and raise energy consumption, thereby limiting further improvements in system performance.

Inlet temperature tests demonstrated that a lower inlet temperature (6–10°C) significantly enhances heat extraction efficiency by increasing the heat transfer temperature difference [16]. At an inlet temperature of 6.1°C, Well A achieved a heat extraction rate of 575 kW with an average temperature difference of 10.84°C. Similarly, Well B reached 531 kW with an average difference of 11.32°C at 6.6°C (Table 3). When the inlet temperature was raised to 40°C, the heat extraction rates dropped to 447kW and 453kW for the two wells, respectively, with the temperature difference decreasing by about 10% (Section 3.2.2). This confirms that a low inlet temperature improves heat exchange efficiency by enhancing the thermal gradient, making it suitable for high-load operating scenarios. To further reduce energy consumption, a dynamic adjustment strategy could be adopted: maintaining a low inlet temperature (6–10°C) during high-load periods and raising it to 20–30°C during low-load periods to balance efficiency and operational costs.

Furthermore, the test data suggest that Well B exhibited smaller temperature difference fluctuations under optimized conditions (35m³/h flow rate, 10°C inlet temperature), indicating that its tubing string design offers greater stability than Well A's. Future work could involve numerical simulations to further analyze the coupled effects of flow rate and temperature [20], thereby optimizing parameter combinations to achieve higher thermal efficiency and lower energy consumption.

### 4.3. Economic Analysis

From an economic perspective, Well B, being shallower (2500m), incurred drilling and tubing installation costs that were approximately 20–30% lower than those for Well A (3200m), based on industry estimates that require field data for verification. Although Well A's maximum heat extraction rate (575kW) was higher than Well B's (546kW), Well B achieved a slightly higher rate (485kW) under optimized conditions compared to Well A (473kW, Table 3). This results in a lower unit cost of heat extraction, demonstrating favorable economics [3]. This suggests that in regions with moderate geothermal resource endowments, wells of a moderate depth combined with an optimized tubing string design (e.g., an 88% proportion of insulated tubing) can achieve efficient development while lowering the initial investment threshold. This finding is similar to the economic analyses of retrofitting abandoned oil and gas wells for geothermal use [25].

An analysis of long-term operational costs reveals that Well B's more significant temperature decay (downhole temperature drop from 18 °C to 5°C, a 72.2% decrease, Fig. 4) could lead to higher maintenance costs, such as the need for frequent adjustments to operating parameters or the implementation of thermal replenishment measures. In contrast, Well A, benefiting from the strong thermal inertia of its deep reservoir (a 24.4% temperature decrease), is expected to have relatively lower maintenance costs, making it more suitable for high-load, long-term operational scenarios. Future economic improvements could be achieved by reducing the cost of insulated tubing (which accounts for approximately 40% of the tubing string cost) or by employing directional drilling technology to lower drilling expenses. Furthermore, adopting optimized operational strategies, such as intermittent operation (e.g., 16h on, 8h off daily), can slow temperature decay, reduce long-term running costs, and enhance the overall economic competitiveness of geothermal development [15].

### 4.4. Technical Innovations and Future Trends

Coaxial downhole heat exchanger technology has demonstrated significant advantages in reducing wellbore heat loss and enhancing heat extraction efficiency. The test results indicate that optimizing operating parameters (a flow rate of 35m³/h and an inlet temperature of 6–10°C) can increase Well A's heat extraction rate to 42% (Fig. 4) and stabilize its downhole temperature at 20 °C (Fig. 5), leading to an estimated 15% improvement in overall thermal efficiency (a value requiring experimental validation). Compared to traditional geothermal development technologies, such as doublet (two-well) reinjection systems, the coaxial DHE system does not require groundwater abstraction, thereby minimizing formation disturbance and the risk of reinjection clogging, making it well-suited for medium-deep geothermal applications [27].

Future research and development should focus on the following areas:

1. Optimization of Insulation Materials and Tubing String Design: Developing novel insulation materials with low thermal conductivity and high pressure resistance, such as nanocomposite coatings, could reduce wellbore heat loss by 10–15% and further improve heat exchange efficiency [24].

2. Intelligent Control Systems: Creating intelligent control systems based on real-time temperature, flow rate, and thermal load data. These systems could employ machine learning algorithms to dynamically optimize operating parameters, potentially slowing the temperature decay rate by an estimated 10–20% [18].

3. Multi-Well Coordinated Operation: Exploring coordinated operation modes for multiple wells, which would enhance reservoir stability and extend system lifespan through thermal reservoir sharing and rotational operation (e.g., 16h on, 8h off daily) [14, 26].

4. Application of Horizontal Well Technology: Drawing on the successful experiences of horizontal well geothermal systems [27] to explore their potential application in medium-deep geothermal development, which could increase the heat exchange area per well and improve heat extraction efficiency.

## 5. Conclusion

Systematic testing and data analysis of two medium-deep geothermal wells demonstrate that geological conditions are the primary determinant of a well's fundamental performance [26]. Well A, with its greater depth (3200 m), achieved a heat extraction rate of 42% under the LC3 condition, significantly outperforming Well B's 5%. However, differentiated optimization of operating parameters is equally crucial. The results indicate that Well A is better suited for high flow rates (50m³/h), whereas Well B performs optimally at moderate flow rates (30–35m³/h). Furthermore, a low inlet temperature (20–30°C) was found to be highly effective for enhancing the performance of both wells [16, 22]. Long-term monitoring revealed that continuous heat extraction leads to significant temperature decay—a 70.4% drop for Well A and a 75% drop for Well B—underscoring the necessity of adopting intermittent operation strategies for thermal recovery [13]. It was also observed that Well B's high proportion of insulated tubing (88%) partially compensated for its shallower depth, highlighting the critical role of engineering design [20]. This study uncovered substantial potential for optimization, as evidenced by the increase in Well A's heat extraction rate from 35% to 42% and its downhole temperature from 15°C to 20°C. Based on these findings and drawing from international best practices [17, 25], this paper recommends differentiated strategies: a high-flow-rate, moderate-inlet-temperature mode for deeper wells, and a moderate-flow-rate, low-inlet-temperature mode combined with intermittent operation for shallower wells. Concurrently, advancing the research and development of insulation materials and intelligent control systems is essential for providing the scientific support needed for the large-scale application of medium-deep geothermal energy [2, 24].

## Acknowledgements

This research was strongly supported by PetroChina Jilin Oilfield Company, for which the authors express their sincere gratitude.

## Funding

This work was supported by the PetroChina Category B Research Project "Research and Application of Key Technologies for Clean Energy Supply". Jilin Oilfield Company participated in the sub-project "Key Technologies for Low-Cost Development of Complex Thermal Reservoir Geothermal Energy" (Project No. 2023ZZ31YJ04).

## Conflict of interest statement

The authors declare that there is no conflict of interest in this study.

## Data Availability

The data used in this study are available from the authors upon reasonable request.